\documentclass[11pt]{iopart}

\usepackage{iopams} 
\usepackage{mathrsfs}
\usepackage{graphicx}
\usepackage{hyperref}
\usepackage{color}

\eqnobysec
\definecolor{CiteColor}{rgb}{0,0,0.35}
\definecolor{URLColor}{rgb}{0,0,0.35}
\hypersetup{colorlinks,citecolor=CiteColor,urlcolor=URLColor}

\newcommand{\beq}{\begin{equation}}
\newcommand{\eeq}{\end{equation}}
\newcommand{\ud}{\mathrm{d}}
\newcommand{\calH}{\mathcal{H}}

\newcommand{\av}[1]{\langle #1 \rangle}

\begin{document}

\title{A Note on Celestial Mechanics in Kerr Spacetime}

\author{Alexandre Le Tiec}

\ead{letiec@obspm.fr}

\address{Laboratoire Univers et Th\'eories (LUTh), Observatoire de Paris, CNRS, \\ Universit\'e Paris Diderot, 5 place Jules Janssen, 92190 Meudon, France}

\begin{abstract}
The Hamilton-Jacobi equation for test particles in the Kerr geometry is separable. Using action-angle variables, we establish several relations between various physical quantities that characterize bound timelike geodesic orbits around a spinning black hole, including the particle's rest mass, energy, angular momentum, mean redshift and fundamental frequencies. These relations are explicitly checked to hold true in the particular case of equatorial circular orbits. An application to the gravitational wave-driven, adiabatic inspiral of extreme-mass-ratio compact binaries is briefly discussed.
\end{abstract}


\maketitle

\section{Introduction}

Ever since Carter's unexpected proof of the separability of the Hamilton-Jacobi equation for test particles orbiting spinning black holes \cite{Ca.68}, timelike geodesics of the Kerr geometry have been explored in great detail, \textit{e.g.} \cite{Ba.al.72,Wi.72,Cha,StTs.87,Bi.al.93,Mi.03,DrHu.04,LePe.08,FuHi.09,Wa.al.13}. In particular, several authors have studied the Hamiltonian mechanics of test masses in the Kerr spacetime while employing action-angle variables \cite{Sc.02,HiFl.08}, a class of canonical variables well adapted to the description of completely integrable dynamical systems \cite{LaLi2,Arn}. In this paper, by making use of such variables, we establish some previously overlooked relationships between various physical quantities characterizing bound orbits around a spinning black hole, including the particle's rest mass, energy, axial angular momentum, mean redshift, and fundamental frequencies. These relationships are somewhat reminiscent of the first law of black hole mechanics \cite{Ba.al.73}, Smarr's formula \cite{Sm.73}, and their generalizations to binary systems of compact objects \cite{Fr.al.02,Le.al.12,Bl.al.13,GrLe.13}.

The remainder of this paper is organized as follows: After briefly reviewing some well-known results on the Hamiltonian mechanics of test particles in the Kerr geometry in section \ref{sec:HamiltonKerr}, we establish ``first law'' type relations and the associated first integrals in the next section \ref{sec:NewRelations}. These new results are checked to hold, in section \ref{sec:verif}, in the particular case of equatorial circular orbits. Finally, we discuss an application to the gravitational wave-driven, adiabatic inspiral of extreme-mass-ratio compact binaries in section \ref{sec:adiabatic}. Our conventions are those of Ref.~\cite{Wal}. In particular, the metric signature is $+2$ and we use ``geometrized units'' where $G = c = 1$. Latin indices $a,b,\dots$ are abstract, while Greek indices $\mu,\nu,\dots$ are used for coordinate components in a particular coordinate system.	

\section{Hamiltonian mechanics in Kerr spacetime}\label{sec:HamiltonKerr}

We consider a test particle of rest mass $\mu$ on a timelike geodesic $\gamma$ of the Kerr geometry $g_{ab}(x;M,S)$ of mass $M$ and angular momentum $S$. We denote the timelike Killing field (normalized to $-1$ at infinity) by $t^a$ and the axial Killing field (with integral curves of parameter length $2\pi$) by $\phi^a$. In the absence of an electromagnetic field, a Hamiltonian that generates geodesic motion is \cite{Ca.68}
\beq\label{H}
	H(y,p;M,S) = \frac{1}{2} \, g^{ab}(y;M,S) \, p_a p_b \, ,
\eeq
where $y$ and $p$ are the particle's canonical position and four-momentum. (Hereafter, we shall omit the dependence of $H$ on the black hole parameters $M$ and $S$.) Using $\lambda \equiv \tau / \mu$ as an affine parameter, where $\tau$ is the proper time elapsed along $\gamma$, Hamilton's canonical equations of motion read
\beq\label{eom}
	\frac{\ud y^\mu}{\ud \lambda} = \frac{\partial H}{\partial p_\mu} \, , \quad \frac{\ud p_\mu}{\ud \lambda} = - \frac{\partial H}{\partial y^\mu} \, .
\eeq
In particular, we have the usual relationship $p_a = \mu \, u_a$ between the four-momentum $p_a$ and the unit four-vector $u^\mu = \ud y^\mu / \ud \tau$ tangent to $\gamma$, such that $H = - \frac{1}{2} \, \mu^2$ ``on shell.''

The 8-dimensional dynamical system \eref{eom} has 4 first integrals: the particle's energy and axial component of the orbital angular momentum, $E = - t^a p_a$ and $L_z = \phi^a p_a$, the Carter constant $Q = K^{ab} p_a p_b$ associated with the Killing tensor $K_{ab}$ of the Kerr geometry \cite{Ca.68}, and the Hamiltonian $H = - \frac{1}{2} \, \mu^2$ itself. The 4 first integrals $P_\alpha \equiv (H,E,L_z,Q)$ are independent (for non-degenerate orbits) and in involution, \textit{i.e.}, have vanishing Poisson brackets \cite{HiFl.08}. Hence, the dynamical system is completely integrable. For bound orbits, a generalization of the Liouville-Arnol'd theorem for dynamical systems with non-compact level sets ensures the existence of generalized \textit{action-angle variables} \cite{Fi.al.03}. Then, using the complete solution of the Hamilton-Jacobi equation in Boyer-Lindquist (BL) coordinates $(t,r,\theta,\phi)$, Hamilton's characteristic function can be used to perform a Type II canonical transformation with a time-independent generating function, yielding a new Hamilonian $\calH(q_\alpha,J_\alpha) = H(y^\mu,p_\mu)$ expressed in generalized action-angle variables $q_\alpha,J_\alpha$. The actions $J_\alpha$ are known functions of the first integrals $P_\alpha$ only; in particular $J_t = - E$ and $J_\phi = L_z$. We refer the reader to Refs.~\cite{Sc.02,HiFl.08} for a detailed account of that construction. In terms of action-angle variables, the canonical equations of motion take the simple form
\beq\label{EOM}
	\frac{\ud q_\alpha}{\ud \lambda} = \frac{\partial \calH}{\partial J_\alpha} \equiv \Omega_\alpha \, , \quad \frac{\ud J_\alpha}{\ud \lambda} = - \frac{\partial \calH}{\partial q_\alpha} = 0 \, .
\eeq
Since the Hamiltonian $\calH(J_\alpha)$ does not depend on the generalized angles $q_\alpha$, the actions $J_\alpha$ are constants of the motion. This, in turn, implies that the \textit{fundamental frequencies} $\Omega_\alpha$ are also constants of the motion. The wide class of coordinate transformations that leave the actions $J_\alpha$ and frequencies $\Omega_\alpha$ unchanged is discussed extensively in Ref.~\cite{HiFl.08}. Schmidt \cite{Sc.02} (see also Ref.~\cite{FuHi.09}) provides explicit expressions, in terms of Legendre and complete elliptic integrals, for the angular frequencies $\omega_\alpha \equiv \Omega_\alpha / \Omega_t$ as functions of $P_\alpha$. Interestingly, while the first integrals uniquely specify an orbit (up to initial conditions), the fundamental frequencies $(\omega_r,\omega_\theta,\omega_\phi)$ do not \cite{Wa.al.13}.

\section{Variational first laws and first integral relations}\label{sec:NewRelations}

For two neighboring solutions of the Hamiltonian dynamics \eref{EOM}, a general variation of the Hamiltonian $\calH(J_\alpha;M,S)$ immediately gives
\beq\label{deltaH}
	\delta \calH = \sum_\alpha \Omega_\alpha \, \delta J_\alpha + \partial_M \calH \, \delta M + \partial_S \calH \, \delta S \, .
\eeq
Here, the partial derivatives of the Hamiltonian with respect to the black hole mass and spin are computed while holding the canonical variables fixed. Notice that all terms in Eq.~\eref{deltaH} are constant, because $\calH$ is a function of the constants of the motion $J_\alpha$ and of the black hole parameters $(M,S)$. Then, using the equalities $J_t = - E$ and $J_\phi = L_z$, as well as the ``on shell'' constraint $\calH = - \frac{1}{2} \, \mu^2$, we obtain the variational relationship
\beq\label{deltaE}
	\delta E = \omega_r \, \delta J_r + \omega_\theta \, \delta J_\theta + \omega_\phi \, \delta L_z + \av{z} \, \delta \mu \nonumber + \frac{\av{z}}{\mu} \left( \partial_M \calH \, \delta M + \partial_S \calH \, \delta S \right) .
\eeq
The fundamental frequencies $\omega_\alpha = \Omega_\alpha / \Omega_t$ are defined with respect to the proper time $t$ of an asymptotically far static observer, and $\av{z} \equiv \av{\ud \tau / \ud t} =  \mu / \Omega_t$ is the phase-space averaged redshift factor. Equation \eref{deltaE} holds for \textit{any} two neighboring bound timelike geodesic orbits in \textit{any} two neighboring Kerr black hole spacetimes. It is reminiscent of the first law of black hole mechanics \cite{Ba.al.73}, and even more of its generalizations to binary systems of compact objects \cite{Fr.al.02,Le.al.12,Bl.al.13,GrLe.13}. Since the Einstein equation does not contain any privileged mass scale, the particle's energy $E$ must be a homogeneous function of degree one in the variables $(J_r^{1/2},J_\theta^{1/2},L_z^{1/2},m,M,S^{1/2})$. Hence, by application of Euler's theorem together with the variational law \eref{deltaE}, we obtain the first integral relation
\beq\label{E}
	E = 2 \left( \omega_r J_r + \omega_\theta J_\theta + \omega_\phi L_z \right) + \mu \av{z} + \frac{\av{z}}{\mu} \left( M \, \partial_M \calH + 2 S \, \partial_S \calH \right) ,
\eeq
which holds for \textit{any} bound (timelike geodesic) orbit around \textit{any} Kerr black hole. This algebraic relation is reminiscent of Smarr's formula \cite{Sm.73} or similar expressions valid for binary systems of compact objects \cite{Le.al.12,Bl.al.13,GrLe.13}.

Since the action variables $E$, $J_r$, $J_\theta$ and $L_z$ are linear in the momentum $p_a = \mu \, u_a$, they must be proportional to the mass $\mu$ of the particle. Thus, we may write $E \equiv \mu \, e$, $J_r \equiv \mu \, j_r$, $J_\theta \equiv \mu \, j_\theta$ and $L_z \equiv \mu \, l_z$, where the specific variables $e$, $j_r$, $j_\theta$ and $l_z$ do not depend on $\mu$. Therefore, plugging these expressions into Eq.~\eref{deltaE}, and noticing that the variations $\delta M$, $\delta S$ and $\delta \mu$ are independent, the coefficient multiplying $\delta \mu$ must vanish. This yields the simple algebraic formula
\beq\label{e}
	e = \omega_r j_r + \omega_\theta j_\theta + \omega_\phi l_z + \av{z} \, ,
\eeq
which must also be valid for any bound orbit around any Kerr black hole. Heuristically, this expression shows how the energy per unit mass $e$ of the particle is equally split into radial, polar, azimuthal, and ``temporal'' components. On the other hand, in terms of specific variables, the particle Hamiltonian first law \eref{deltaE} reduces to
\beq\label{deltae}
	\delta e = \omega_r \, \delta j_r + \omega_\theta \, \delta j_\theta + \omega_\phi \, \delta l_z + \av{z} \left( \partial_M \bar{\calH} \, \delta M + \partial_S \bar{\calH} \, \delta S \right) ,
\eeq
where we introduced the dimensionless Hamiltonian $\bar{\calH} \equiv \calH / \mu^2$. There is no first integral relation associated with this variational law, because the energy per unit mass $e$ is \textit{not} a homogeneous function of degree one in the variables $(j_r^{1/2},j_\theta^{1/2},l_z^{1/2},M,S^{1/2})$, since the additional mass scale $\mu$ enters into the problem.

\section{Verification for circular equatorial orbits}\label{sec:verif}

For circular equatorial orbits, the radial and polar action variables vanish: $J_r = J_\theta = 0$.\footnote{However, the libration-type frequencies $\omega_r$ and $\omega_\theta$ do not vanish \cite{Sc.02}; they characterize the stability of slightly eccentric and inclined orbits, respectively.} The averaging over phase space of the redshift $z = \ud \tau / \ud t$ is trivial, as it is constant along the orbit. All the physical variables of interest have closed-form expressions in terms of the BL coordinate radius $r$, or equivalently of the azimuthal frequency $\omega_\phi$, namely \cite{Ba.al.72}
\numparts
	\begin{eqnarray}
		e = \frac{1 - 2v^2 + \chi v^3}{\sqrt{1 - 3v^2 + 2 \chi v^3}} \, , \\
		l_z = \frac{M}{v} \, \frac{1 - 2 \chi v^3 + \chi^2 v^4}{\sqrt{1 - 3v^2 + 2 \chi v^3}} \, , \\
		\av{z} = \frac{\sqrt{1 - 3v^2 + 2 \chi v^3}}{1 + \chi v^3} \, ,
	\end{eqnarray}
\endnumparts
where $\chi \equiv S / M^2$ is the dimensionless Kerr parameter, and we introduced the convenient notation $v^2 \equiv M / r = (M \omega_\phi)^{2/3} / (1 - \chi \, M \omega_\phi)^{2/3}$. Similarly, closed-form expressions for the partial derivatives of the Hamiltonian with respect to the black hole mass and spin can easily be derived by using the equality $\partial_{M,S} \bar{\calH} \vert_{q,J} = \partial_{M,S} \bar{H} \vert_{y,p}$ for circular equatorial orbits, and by writing down \eref{H} explicitly, \textit{e.g.}, in BL coordinates; one finds
\numparts
	\begin{eqnarray}
		\partial_M \bar{\calH} = - \frac{v^2}{M} \, \frac{1 + 2 \chi v^3 - \chi^2 v^4}{1 - 3v^2 + 2 \chi v^3} \, , \\
		\partial_S \bar{\calH} = \frac{v^5}{M^2} \, \frac{2 - \chi v}{1 - 3v^2 + 2 \chi v^3} \, .
	\end{eqnarray}
\endnumparts
Using the above expressions, one can easily check that Eqs.~\eref{deltaE}--\eref{deltae} are satisfied. The particle Hamiltonian first law \eref{deltaE} and first integral relation \eref{E} were key ingredients in the recent derivation of the horizon angular velocity and surface gravity of a spinning black hole perturbed by a small corotating moon \cite{GrLe.13}. Although the relations \eref{deltaE}--\eref{deltae} cannot be checked analytically for generic orbits, they could be checked numerically for any given non-circular, non-equatorial bound timelike geodesic of the Kerr geometry.\footnote{I am grateful to S. Isoyama for letting me know of an alternative proof of (3.4) \cite{Is.al.14} and to S. Hughes for his observation that this relation is satisfied for every bound timelike Kerr geodesic that he checked.}

\section{Adiabatic inspiral of extreme-mass-ratio binaries}\label{sec:adiabatic}

We now go beyond the test-particle approximation $\mu \to 0$, including the leading effects of gravitational radiation-reaction on the dynamics, such that the orbital parameters evolve in time. In the extreme-mass-ratio limit $\mu \ll M$, any change in the particle's rest mass $\mu$ can be neglected during the entire orbital evolution. Moreover, the two-timescale analysis of Hinderer and Flanagan \cite{HiFl.08} shows that, in the \textit{adiabatic approximation}, the black hole mass $M$ and spin $S$ remain constant too; their time evolution, which is driven by the absorption of gravitational radiation, is a post-1-adiabatic effect. Hence, in the adiabatic approximation, the relationship \eref{deltaE} with $\delta \mu = \delta M = \delta S = 0$ shows that the instantaneous rates of change of the actions $E(t)$, $J_r(t)$, $J_\theta(t)$ and $L_z(t)$ are related by
\beq\label{<Edot>}
	\dot{E} = \omega_r \, \dot{J}_r + \omega_\theta \, \dot{J}_\theta + \omega_\phi \, \dot{L}_z \, ,
\eeq
where the overdot stands for a derivative with respect to coordinate time $t$. Generically, the orbital motion is ergodic, ensuring that the averaging over phase space is equivalent to an infinite time average.\footnote{With the exception of ``resonant orbits,'' \textit{i.e.}, orbits for which the libration frequencies $\omega_r$ and $\omega_\theta$ are commensurable \cite{FlHi.12}. However, these correspond to a subset of measure zero in the set of bound orbits.} Therefore, by averaging Eq.~\eref{<Edot>} over a time interval that is large compared to the orbital timescales, but small compared to the radiation-reaction timescale \cite{Dr.al.05,Sa.al.06}, and using the conservation laws $\av{\dot{E}} = - \mathcal{F}$ and $\av{\dot{L}_z} = - \mathcal{G}$, where $\mathcal{F}$ and $\mathcal{G}$ are the gravitational-wave fluxes of energy and angular momentum to infinity and down the event horizon \cite{Cu.al.94,Sh.94}, we immediately get
\beq\label{fluxes}
	\omega_r \, \av{\dot{J}_r} + \omega_\theta \, \av{\dot{J}_\theta} = - \mathcal{F} + \omega_\phi \, \mathcal{G} \, .
\eeq
For any given bound timelike geodesic orbit, the gravitational-wave fluxes $\mathcal{F}$ and $\mathcal{G}$ can be computed numerically, with high accuracy, using either the Teukolsky equation \cite{Te.72} or the Sasaki-Nakamura equation \cite{SaNa.82}. The relation \eref{fluxes} can thus be used to characterize the orbital evolution of extreme-mass-ratio inspirals (EMRIs) in the radial and/or polar directions, providing quantitative information on the leading-order (adiabatic) evolution of the eccentricity and/or inclination of the orbit. For instance, the radial action $J_r$ of a generic equatorial orbit ($J_\theta = 0$) must decrease (on average) if $\mathcal{F} > \omega_\phi \, \mathcal{G}$, and increase otherwise. This could be used to gain some insight on the eccentricity increase observed just before the transition from inspiral to plunge \cite{Ap.al.93,Ta.al.93,Cu.al.94,Ke.98,KoKe.02}.

\ack

It is a pleasure to thank {\'E}. {\'E}. Flanagan, T. Hinderer, S. Isoyama and T. Tanaka for a careful reading of this manuscript and for useful comments.

\section*{References}

\bibliographystyle{iopart-num}
\bibliography{}

\end{document}